\PassOptionsToPackage{table,xcdraw}{xcolor}
\documentclass[sigconf]{acmart}

\AtBeginDocument{%
  }

\copyrightyear{2024}
\acmYear{2024}
\setcopyright{rightsretained}
\acmConference[ASEW '24]{39th IEEE/ACM International Conference on
Automated Software Engineering Workshops}{October 27-November 1,
2024}{Sacramento, CA, USA}
\acmBooktitle{39th IEEE/ACM International Conference on Automated Software
Engineering Workshops (ASEW '24), October 27-November 1, 2024, Sacramento,
CA, USA}
\acmDOI{10.1145/3691621.3694953}
\acmISBN{979-8-4007-1249-4/24/10}

\usepackage[table,xcdraw]{xcolor}
\usepackage[framemethod=TikZ]{mdframed}
\usepackage{float}
\usepackage{subcaption}
\usepackage{listings}
\usepackage{adjustbox}
\definecolor{cyan1}{RGB}{0,255,255}
\definecolor{codegreen}{rgb}{0,0.6,0}
\definecolor{codegray}{rgb}{0.5,0.5,0.5}
\definecolor{codepurple}{rgb}{0.58,0,0.82}
\definecolor{backcolour}{rgb}{0.95,0.95,0.92}

\usepackage{rotating}
\usepackage{todonotes}
\definecolor{lightblue}{RGB}{200, 220, 255}

\lstdefinestyle{mystyle}{   
    commentstyle=\color{codegreen},
    keywordstyle=\color{magenta},
    numberstyle=\tiny\color{codegray},
    stringstyle=\color{codepurple},
    basicstyle=\ttfamily\footnotesize,
    breakatwhitespace=false,         
    breaklines=true,                 
    captionpos=b,   
    frame=single,
    keepspaces=true,                 
    numbers=left,                    
    numbersep=5pt,                  
    showspaces=false,                
    showstringspaces=false,
    showtabs=false,                  
    tabsize=2
}

\begin{document}

\title{Advancing Android Privacy Assessments with Automation}

\author{Mugdha Khedkar}
\affiliation{%
  \institution{\textit{Heinz Nixdorf Institute \\ Paderborn University}}
  \city{Paderborn}
  \country{Germany}
}
\email{mugdha.khedkar@uni-paderborn.de}

\author{Michael Schlichtig}
\affiliation{%
  \institution{\textit{Heinz Nixdorf Institute \\ Paderborn University}}
  \city{Paderborn}
  \country{Germany}
}
\email{michael.schlichtig@uni-paderborn.de}

\author{Eric Bodden}
\affiliation{%
  \institution{\textit{Heinz Nixdorf Institute \\ Paderborn University and Fraunhofer IEM}}
  \city{Paderborn}
  \country{Germany}
}
\email{eric.bodden@uni-paderborn.de}



\begin{abstract}
Android apps collecting data from users must comply with legal frameworks to ensure data protection.
This requirement has become even more important since the implementation of the General Data Protection Regulation (GDPR) by the European Union in 2018.
Moreover, with the proposed Cyber Resilience Act on the horizon, stakeholders will soon need to assess software against even more stringent security and privacy standards. 
Effective privacy assessments require collaboration among groups with diverse expertise to function effectively as a cohesive unit.
  
This paper motivates the need for an automated approach that enhances understanding of data protection in Android apps and improves communication between the various parties involved in privacy assessments. 
We propose the \emph{Assessor View}, a tool designed to bridge the knowledge gap between these parties, facilitating more effective privacy assessments of Android applications.
\end{abstract}

\ccsdesc[500]{Security and privacy~Social network security and privacy}
\ccsdesc[300]{Software and its engineering~Software maintenance tools}
\keywords{automated GDPR compliance, privacy assessment, static analysis}


\maketitle

\section{Introduction}

Any software that reaches the European market needs to adhere to the General Data Protection Regulation (GDPR)~\cite{gdpr}. 
Moreover, with the European Union's Cyber Resilience Act (CRA)~\cite{cra} on the horizon, software developers will soon face the challenge of writing code that complies with even more stringent security and privacy standards. 
These regulations extend to Android applications that gather data from users within the European Union. 

The GDPR defines personal data as \textit{``any information relating to an identified or identifiable natural person, a data subject"}, 
and imposes several obligations on the access, storage and processing of such data. 
Under the GDPR, data protection violations can result in severe financial penalties~\cite{penalties}. 
If these violations cause vulnerabilities and data leaks, additional and similarly severe fines may be levied under the CRA. 


Advocating for transparency, Article~13~\cite{gdprarticles} of the GDPR mandates that the app publishers disclose the collection and processing of personal data to the user by providing documents such as privacy policies. 
However, these privacy policies are often very long and vague, and may not even be authored by the app developers themselves. 
Several studies~\cite{privacypolicytrust,automatedriskanalysis,guileak,ppviolationappcode,PTPDroid} have consistently shown significant discrepancies between privacy policies and the actual source code, undermining their accuracy and misleading users.

To address these inaccuracies, Google launched the data safety section~\cite{data} in 2022, shifting the responsibility of privacy-related reporting to app developers. 
This necessitates the completion of a form on Google's Play Console, outlining how apps collect, share, and secure user data. 

Google's data safety section (DSS) form consists of three main sections: \textit{data sharing, data collection, and security practices}. 
User data is categorized into different types within these sections. 
Before creating a store listing for an Android app, developers must complete this DSS form. 
This requires manual effort which may lead to inaccuracies in reporting~\cite{datalabels}, providing users with a false sense of privacy. 
In 2022, Google introduced Checks~\cite{googlechecks}, a paid service that assists app developers with privacy compliance, and completing the data safety section. 
However, its cost could hinder startups or independent app developers from using its services.
Recently, open-source alternatives~\cite{matcha,privadoai} have been introduced to assist developers in accurately completing the DSS form. 

While accurately filling out the DSS form is the developers' responsibility, ensuring GDPR compliance through assessments requires collaboration among groups with a diverse expertise. 
Article~35~\cite{gdprarticles} of the GDPR describes the Data Protection Impact Assessment (DPIA)~\cite{dpia} as \textit{``an assessment of the impact of the envisaged processing operations on the protection of personal data.''}. 
Conducting the DPIA is a responsibility of the data controller (software providers or organizations) and requires oversight from the Data Protection Officer (DPO), a legal expert appointed by the data controller. 
The DPIA involves a systematic analysis of software to identify and mitigate data protection risks, requiring collaboration between DPOs, legal experts, and technical experts (such as app developers). 
Controllers are obligated to document the DPIA results and share them with the European Data Protection Board~\cite{edpb}. 

The documents discussed thus far—the privacy policy, DSS form, and DPIA documentation—are authored by and directed towards different target groups, and nonetheless must be consistent with one another, and with the source code. 
Given the diverse expertise required to ensure accurate privacy assessments, we ask: \textit{How can one bridge the knowledge gap between app developers, Data Protection Officers (privacy specialists), and legal experts to ensure seamless privacy assessments?}

In this paper, we propose the \emph{Assessor View}, a tool designed to address this knowledge gap among the various parties involved in privacy assessments. 
It visualizes source code components in terms of a data privacy vocabulary~\cite{dpv2}, and provides multiple views with different levels of granularity and types of visualization tailored for legal and technical experts, respectively.
Throughout this paper, we use the term \textit{privacy assessors} to collectively refer to DPOs and legal experts. 
\section{Problem}
\label{problem}

Consider Alice, the newly appointed Data Protection Officer of a company, who initiates a Data Protection Impact Assessment. 
To complete the assessment, she requires app developers' assistance in answering the DPIA questions. 
She seeks help of Bob, a Java programmer unfamiliar with legal privacy frameworks. 
They use the sample DPIA template~\cite{dpiatemplate} and aim to address questions regarding the source of the data collected; the use, storage, and deletion of data; its processing and sharing; and whether the data is pseudonymized. 

Without tool support, Alice and Bob must manually sift through extensive lines of code to find answers. While Bob navigates the code proficiently, Alice struggles due to her limited programming expertise. Conversely, Alice provides legal insights to Bob, but she struggles to correlate legal terms with source code components. 
This leads to an extensive discussion with the legal team, during which Bob must convey technical concepts to non-technical experts and persuade them regarding the implementation of data protection measures. After weeks of manual labor and discussion, they eventually complete the assessment successfully.

Conducting a DPIA involves identifying and mitigating data protection risks, requiring collaboration between developers, DPOs, and legal experts. App developers may often spend considerable time convincing the team of appropriate data protection measures. 

Alice and Bob could have been better supported with a tool to systematically assess privacy-by-design concepts in source code. 
This highlights the need to design and implement an automated approach to bridge the knowledge gap between app developers and privacy assessors
, ensuring seamless privacy assessments. 

While existing research provides foundational knowledge to address this issue and better support developers and legal experts in conducting the DPIA, several challenges still need to be addressed.  

Pandit et al.~\cite{dpvreport} identified the absence of standard vocabularies to describe personal data, data handling purposes, and categories of processing. 
They introduced the data privacy vocabulary (DPV), which describes various components and relationships between them~\cite{dpv2}. 
While DPV offers a clear ontology for legal experts, mapping source code components to this vocabulary is challenging but crucial for improving communication in privacy assessments. 

Existing static analysis tools~\cite{flowdroid, PTPDroid,atpchecker} aid app developers in assessing whether their app source code and privacy policies fulfil data protection requirements. 
However, these tools are primatily designed for developers. 

Nachtigall et al.'s study on usability criteria of such static analysis tools~\cite{usability} showed that usibility criteria are often only partially addressed or even neglegted. 
Further, static analysis tools that provide a dedicated user interfaces generally fulfil more usability criteria. 
This suggests that tools supporting the DPIA process should incorporate user interfaces to effectively assist both developers and legal experts. 

While significant efforts have been made to assist developers, there is limited research focused on bridging the knowledge gap between developers and legal experts. 
Feiyang Tang et al.'s analysis~\cite{privflow} was among the first to address this gap, aiming to assist app developers and DPOs in evaluating privacy compliance and simplifying the DPIA process. 
They introduced an automatic software analysis technique that presents the results as a graph of privacy flows and operations. 
While their approach aligns with our objectives, we build on our previous work~\cite{mobilesoftnier} which proposed using the Data Privacy Vocabulary (DPV) to make the analysis results more understandable to legal experts.

\section{Approach}
\label{workflow}

\begin{figure}[t]
\begin{center}
\includegraphics[width=0.45\textwidth]{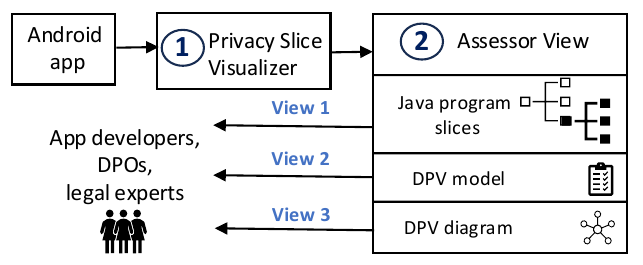}
\caption{Our approach. Contributions of Module 1 are under submission.}
\label{fig:workflow}
\end{center}
\end{figure}

Our approach to address the problem discussed in Section~\ref{problem} comprises two key components: Privacy Slice Visualizer (Module~\textcircled{1} in Figure~\ref{fig:workflow}), and the Assessor View (Module~\textcircled{2} in Figure~\ref{fig:workflow}). 
Initially, the \textcircled{1} \textit{Privacy Slice Visualizer} accepts an APK as input, statically slices the source code from privacy-related data sources, and visualizes the results. Subsequently, the ~\textcircled{2} \textit{Assessor View} accepts these program slices as input and and translates them into a graphical representation familiar to the privacy assessors, enhancing their comprehension of the code. 

The \textbf{Privacy Slice Visualizer} (Module~\textcircled{1} in Figure~\ref{fig:workflow}) uses static analysis to identify privacy-related data sources and then slices the source code to retain the program statements affected by the data source. It is an extension of Jicer~\cite{jicer}, a static program slicer that works with Jimple, an intermediate representation of Java code~\cite{soot}. 
The resulting program slices offer a graphical depiction of control and data dependencies within the source code, represented in Jimple. 
The contributions of Privacy Slice Visualizer are under submission. 
Pilot studies conducted to assess the usability of these program slices have confirmed the need for a simplified view that will facilitate stakeholders' comprehension of the source code.

To address this, the \textbf{Assessor View} (Module~\textcircled{2} in Figure~\ref{fig:workflow}) will convert Jimple program slices into views with different abstraction levels. 
These views will map properties observed within the program slices to specific privacy-related observations. 

Such mapping, developed through extensive discussions with DPOs, aims to flag warnings and suggest data minimization opportunities, providing a higher abstraction level to explain the code to privacy assessors.

We plan to implement three views: \textcolor[HTML]{4472C4}{\textbf{View 1}} with Java (source code level) program slices for app developers, \textcolor[HTML]{4472C4}{\textbf{View 2}} with the intermediary level DPV model for the DPOs, and \textcolor[HTML]{4472C4}{\textbf{View 3}} with the high level DPV diagram for the legal experts.

\begin{figure}[t]
    \centering
\includegraphics[width=0.45\textwidth]{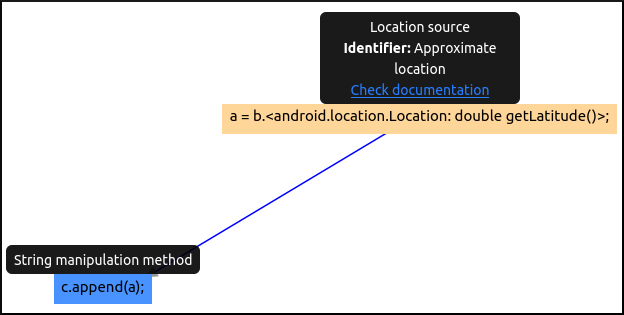}
\caption{\textcolor[HTML]{4472C4}{\textbf{View 1}} (Java slice) for Roidsec}
\label{fig:roidsecjavaslice}
\end{figure}

Figure~\ref{fig:roidsecjavaslice} illustrates a mock-up of \textcolor[HTML]{4472C4}{\textbf{View 1}} (Java program slice) from Roidsec\footnote{\url{https://github.com/TaintBench/roidsec}}, a real-world app from the TaintBench suite~\cite{taintbench}. 
This view will allow assessors to examine the app at source code level. 
In this example, \textit{latitude} (labeled as location data) is collected and appended to a string variable \textit{c} but does not seem to be further used in the source code. 
Our examination shows that this is the only slice originating from latitude, and hence prompts the question: why was latitude collected in the first place? 
This pattern aligns with the privacy-related behavior of data minimization (Article 5~\cite{gdprarticles}), and may be used to suggest insights to the assessors. 

For a higher-level overview, the next view will represent the slice as a DPV model (\textcolor[HTML]{4472C4}{\textbf{View 2}}), presenting the app in terms of the data privacy vocabulary (DPV)~\cite{dpv2}. 
The DPV provides a detailed, machine-readable taxonomy for privacy-related information, based on GDPR terminology. 
We are collaborating with the DPV maintenance group to enhance this model. 

\begin{figure}[t]
    \centering
\includegraphics[width=0.37\textwidth]{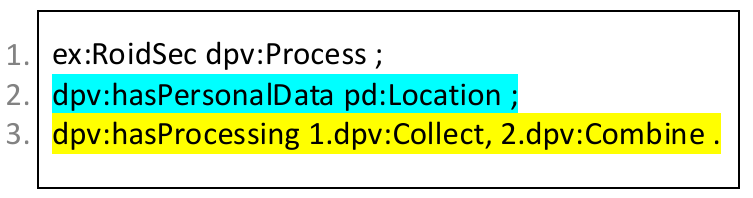}
\caption{\textcolor[HTML]{4472C4}{\textbf{View 2}} (DPV model) for Roidsec; ex = example, dpv = data privacy vocabulary, pd = personal data}
\label{fig:roidsecview2}
\end{figure}

\begin{figure}[t]
    \centering
\includegraphics[width=0.3\textwidth]{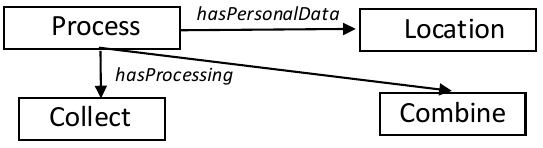}
\caption{\textcolor[HTML]{4472C4}{\textbf{View 3}} (DPV diagram) for Roidsec}
\label{fig:roidsecview3}
\end{figure} 
Figure~\ref{fig:roidsecview2} shows a concrete instance of this model for Roidsec. It uses DPV terms such as personal data categories, and processing categories, with relationships established through predicates like \emph{hasPersonalData} and \emph{hasProcessing}. 
Line 1 in Figure~\ref{fig:roidsecview2} uses the keyword \emph{Process} to define the program slice as a new use case. 
Line 2 suggests that the app collects \colorbox{cyan1}{location}, while line 3 shows that the app first \colorbox{yellow}{collects the data} and then \colorbox{yellow}{combines it} with another variable. 
We will update the model to include control and data dependencies from \textcolor[HTML]{4472C4}{\textbf{View 1}} (cf.~Figure~\ref{fig:roidsecjavaslice}). 
If multiple processing operations are present, we will include their order (as shown in Figure~\ref{fig:roidsecview2}). 

Finally, the Assessor View will convert the DPV model to a concise DPV diagram, summarizing the app at a high level (cf.~Figure~\ref{fig:roidsecview3}, \textcolor[HTML]{4472C4}{\textbf{View 3}}). 
This high-level view will display potential GDPR violations, GDPR compliance warnings, and suggestions to the assessors. 
\textcolor[HTML]{4472C4}{\textbf{View 3}} will highlight the absence of processing operations on collected data and suggest data minimization, prompting app developers to update their source code to only collect data necessary for app functionality. 
It will directly map the violation of the data minimization principle to a \textbf{potential breach} of GDPR Article 5, providing insights to the assessors.



During privacy assessments, assessors can easily switch between different views to ask developers specific questions and understand the technical measures implemented to protect data. 
Alice and Bob can now use the Assessor View for discussions regarding privacy assessments. Alice (DPO) can first examine \textcolor[HTML]{4472C4}{\textbf{View 3}}, which, although concise, lacks detailed information regarding privacy-related data flow. She can then switch to \textcolor[HTML]{4472C4}{\textbf{View 2}} to observe the flow of privacy-related data. If she needs to discuss the intricacies of the source code with Bob (app developer), she can seamlessly switch to \textcolor[HTML]{4472C4}{\textbf{View 1}}, which provides the most detailed information. 

\section{Case Study}
\label{casestudy}

To demonstrate how the Assessor View can assist the privacy assessment team, we manually examined program slices of real-world Android apps generated by the Privacy Slice Visualizer, and observed properties with possible privacy implications.
Additionally, we manually constructed DPV models (\textcolor[HTML]{4472C4}{\textbf{View 2}}) and diagrams (\textcolor[HTML]{4472C4}{\textbf{View 3}}) for these slices to explore the design of the Assessor View.

\begin{table*}[t]
\caption{Properties in program slices and their privacy implications. More examples available at \url{https://zenodo.org/records/13124309}.}
\label{tab:slicesproperties}
\begin{tabular}{lm{5.6cm}ll}
\rowcolor{gray!50}
 \textbf{Program Slice Property}                                                               & \textbf{View 2: DPV Model} & \textbf{View 3: DPV Diagram}   &    \textbf{Privacy Implication}                                                               \\ 
\begin{tabular}[c]{@{}l@{}} Email address collected \\ by Google Play Services \\ (third party library) and \\ written to Google storage. \end{tabular} & 
 {\includegraphics[scale=0.55]{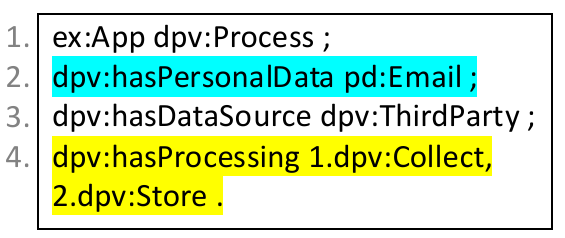}} 
& \adjustbox{valign=c} {\includegraphics[scale=0.5]{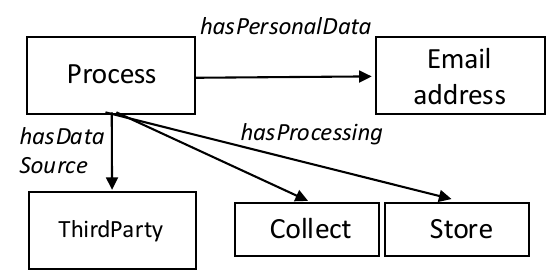}} 
& \begin{tabular}[c]{@{}l@{}} Violation of GDPR \\Article 25. \\ Third party collection \\should adhere to \\GDPR Article 5. \end{tabular}  \\
\rowcolor[HTML]{EDEBEB} 
\begin{tabular}[c]{@{}l@{}} Email address collected, \\ and used to send emails. \end{tabular} & {\includegraphics[scale=0.55]{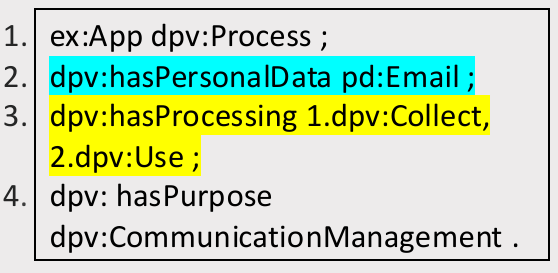}} 
& \adjustbox{valign=c} {\includegraphics[scale=0.5]{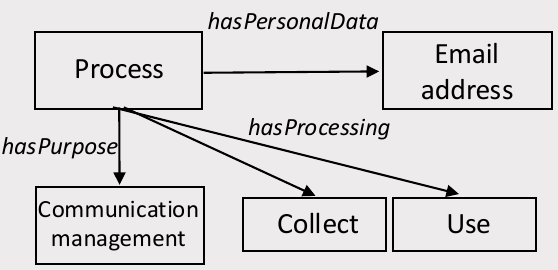}}   & \begin{tabular}[c]{@{}l@{}} Violation of GDPR \\ Article 25. \end{tabular} \\ 
\begin{tabular}[c]{@{}l@{}} Phone number collected, \\ and hashed.   \end{tabular}            &
 {\includegraphics[scale=0.55]{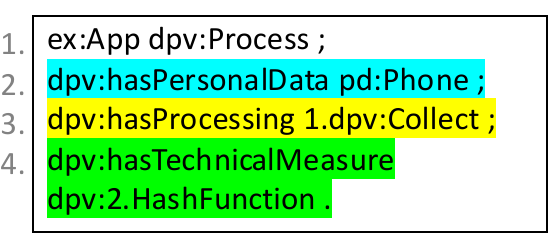}} 
& \adjustbox{valign=c} {\includegraphics[scale=0.55]{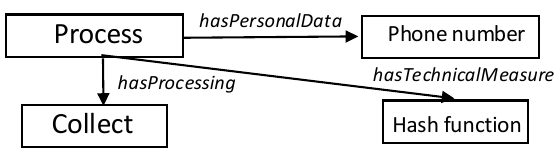}} 
& \begin{tabular}[c]{@{}l@{}} Adherance to GDPR \\Article 25. \end{tabular} \\   
\end{tabular}
\end{table*}

We now examine and discuss some program slices and their properties that could have possible privacy implications (cf.~Table~\ref{tab:slicesproperties}).

The first program slice is from a real-world Android app Steam\footnote{\url{https://play.google.com/store/apps/details?id=com.valvesoftware.android.steam.community&hl=en}}. 
It uses Google Play API methods to collect \colorbox{cyan1}{email address} and then writes it to Google storage. 
According to DPV definitions, this program slice has a \emph{third-party data source} and involves the processing operations of  \colorbox{yellow}{collection} and  \colorbox{yellow}{storage}. 
According to Article 25~\cite{gdprarticles}, personal data should be pseudonymized before being stored or processed. 
\textcolor[HTML]{4472C4}{\textbf{View 2}} and \textcolor[HTML]{4472C4}{\textbf{View 3}} clearly show the absence of pseudonymization before storing the email address, suggesting a \textbf{potential violation} of GDPR Article 25. 
Privacy assessors can thus navigate to \textcolor[HTML]{4472C4}{\textbf{View 1}} (Java slice) to discuss implementation details with the developers, and even use the different views as an evidence of GDPR violation. 
Additionally, the presence of a third-party data source in \textcolor[HTML]{4472C4}{\textbf{View 2}} and \textcolor[HTML]{4472C4}{\textbf{View 3}} indicates that the app must adhere to Chapter 5~\cite{gdprarticles} to ensure lawful third-party data transfer. 

In the next example, the TaintBench app Beita\_com\_beita\_contact~\footnote{\url{https://github.com/TaintBench/beita_com_beita_contact}} collects \colorbox{cyan1}{email address} and uses it to send emails. 
The DPV model records processing operations as \colorbox{yellow}{collection} and  \colorbox{yellow}{usage}, and identifies the purpose of data collection as \emph{communication management} (line 4 in \textcolor[HTML]{4472C4}{\textbf{View 2}}). 
Since neither \textcolor[HTML]{4472C4}{\textbf{View 2}} nor \textcolor[HTML]{4472C4}{\textbf{View 3}} indicates pseudonymization, the app appears to \textbf{violate} Article 25.

The final example is from a TaintBench app Overlay\_android\_samp~\footnote{\url{https://github.com/TaintBench/overlay_android_samp}}, which \colorbox{yellow}{collects} a \colorbox{cyan1}{phone number} and \emph{hashes} it. 
\textcolor[HTML]{4472C4}{\textbf{View 2}} and \textcolor[HTML]{4472C4}{\textbf{View 3}} demonstrate the use of a \colorbox{green}{hash function} as a technical measure before processing the data, thus \textbf{adhering} to Article 25.
\textcolor[HTML]{4472C4}{\textbf{View 2}} and \textcolor[HTML]{4472C4}{\textbf{View 3}} can be used as an evidence of GDPR compliance.

\section{Challenges and Future Plans}
\label{futurework}

We are collaborating with the DPV maintenance group and DPOs to design the mapping between Jimple program slices and the different views, which is crucial for implementing the Assessor View. 
We will conduct studies with app developers and privacy assessors to evaluate our approach and identify areas for improvement. 
We anticipate significant challenges in recruiting assessors for interviews and user studies, and an even greater challenge will be convincing them to trust this automated approach.

While the paper presents simple examples of program slices, most real-world app slices generated by Module 1 are significantly larger (exceeding 100 nodes). 
This complexity highlights the need for manageable high-level views that summarize the source code, making it a strong motivation for our approach. 
However, scalability will be a critical challenge in automatically converting these large slices into different views.
Additionally, maintaining the Assessor View in alignment with the evolving DPV versions will be crucial towards ensuring its continued usability.

Ongoing efforts are required to optimize the static program slicing algorithm in Module 1 and enhance the precision of the resulting program slices. 
While static analysis is useful for creating an initial prototype, future research could build on this foundation by exploring hybrid or dynamic approaches to further streamline privacy assessments. 
\section{Conclusion}
\label{conclusion}

Effective privacy assessments and audits necessitate collaboration among diverse groups with specialized expertise. This paper addresses the research question:
\textit{How can one bridge the knowledge gap between app developers, Data Protection Officers (privacy specialists), and legal experts to ensure seamless privacy assessments?}

In this paper, we have introduced the Assessor View, a tool designed to address this knowledge gap and streamline the privacy assessments of Android applications. 
Tailored to privacy assessors' needs, the envisioned approach will enable quick and effective resolution of inquiries from individuals like Alice and Bob. 
It has the potential to facilitate a cost-effective, transparent, and accurate privacy assessment process.
\begin{acks}
This research is partially funded by the Deutsche Forschungsgemeinschaft (DFG, German Research Foundation) – SFB 1119 – 236615297. 
\end{acks}

\bibliographystyle{ACM-Reference-Format}
\bibliography{sample-base}

\end{document}